\documentclass[useAMS,usenatbib]{mn2e}
\usepackage{graphicx}
\usepackage[usenames]{color}

\title[Pulsations in M dwarf stars]{Pulsations in M dwarf stars}
\author[C. Rodr\'\i guez-L\'opez, J. MacDonald and A. Moya]{C. Rodr\'\i guez-L\'opez$^{1}$\thanks{E-mail:
cristinatrl72@gmail.com (CRL); jimmacd@udel.edu (JM); amoya@cab.inta-csic.es (AM)}, J. MacDonald$^{2}$ and A. Moya$^{3}$\\
$^{1}$Dep. de F\'\i sica Estelar. Instituto de Astrof\'\i sica de Andaluc\'\i a (IAA-CSIC), 18008 Granada, Spain\\
$^{2}$Dep. of Physics and Astronomy, University of Delaware, Newark, DE 19716, USA \\
$^{3}$Dep. de Astrof\'\i sica. Centro de Astrobiolog\'ia (CAB, INTA-CSIC), ESAC Campus, PO Box 78, 28691 Villanueva de la Ca\~nada, Madrid, Spain}

\begin{document}

\date{Accepted 2011 October 11. Received 2011 September 29; in original form 2011 August 04}

\pagerange{\pageref{firstpage}--\pageref{lastpage}} \pubyear{2002}

\maketitle

\label{firstpage}

\begin{abstract}
We present the results of the first theoretical non-radial non-adiabatic pulsational study of M dwarf stellar models with masses in the range 0.1 to 0.5~M$\odot$. We find the fundamental radial mode to be unstable due to an $\epsilon$ mechanism, caused by deuterium (D-) burning for the young 0.1 and 0.2~M$\odot$ models, by non-equilibrium He$^3$ burning for the 0.2 and 0.25~M$\odot$ models of 10$^4$~Myr, and by a flux blocking mechanism for the partially convective 0.4 and 0.5~M$\odot$ models once they reach the age of 500~Myr. The periods of the overstable modes excited by the D-burning are in the range 4.2 to 5.2~h for the 0.1~M$\odot$ models and is of order 8.4~h for the 0.2~M$\odot$ models. The periods of the modes excited by He$^3$ burning and flux blocking are in the range 23 to 40~min. The more massive and oldest models are more promising for the observational detection of pulsations, as their ratio of instability $e$-folding time to age is more favourable.

\end{abstract}


\section{Introduction}
M dwarf stars are cold, low-mass stars on the main sequence that have recently gained attention as being the best candidates to harbour Earth-like planets \citep{delfosse06}. Although they make up about 75\% of the stars in our Galaxy and the solar neighbourhood, their main physical parameters, such as their radii, are poorly understood, and hard to determine observationally due to their low luminosities. 

Evidence has been accumulating for discrepancies between theoretically predicted radii and effective temperatures of low mass stars (LMS), and those derived from observational data. Observational studies of LMS in eclipsing binaries and single LMS have shown that their radii are about 4 to 20\% larger, and their temperatures as much as 5\% lower than predicted by theoretical models (see e.g. \citealt{morales09} and references therein). This discrepancy is usually attributed to effects associated with chromospheric activity or metallicity (see e.g. \citealt{lopez07}).

Another reported problem is age discrepancy: although mean ages derived from lithium isochrones for LMS in young groups agree with those derived from Hertzsprung-Russell diagram (HRD) isochrones, the two methods often give discordant ages when applied to the individual stars in the young groups (see e.g. \citealt{yee10} -YJ10).

Magnetic inhibition of convection provides a possible solution to both problems. This leads to models having larger radii and lower temperatures \citep{mullan01}, and can remove the age discordance \citep{macdonald10}, at least for the stars in the YJ10 sample. A different solution to the radius problem is given by \citet{chabrier07} who reduce the mixing length parameter ($\alpha$) which also gives models with larger radii and lower  temperatures. The two approaches have a different signature on the stellar models: while reducing $\alpha$ affects mainly the outer super-adiabatic regions of a surface convection zone, magnetic inhibition of convection affects the whole of the convection zone. This produces different profiles for the sound speed ($c_s$) in the interior of the star, and raises the possibility of differentiating between the two approaches through asteroseismic techniques. A preliminary calculation \citep{macdonald10} using a magnetic and a non-magnetic stellar model show that the so called large and small frequency separations (related to $c_s$ and its derivative, respectively) given by the asymptotic theory \citep{tassoul80} could distinguish between the two theoretical formulations, reaching differences in the two quantities up to 2 and 30\%, respectively.

Here we present the first non-radial, non-adiabatic pulsational study of M dwarf stellar models. The destabilization of the fundamental radial mode is achieved through an active $\epsilon$ mechanism caused either by deuterium (D-) or He$^3$ nuclear burning. For the more massive and older models considered, the blocking of the radiative flux at the base of the outer convective envelope explains the destabilization.

\section{Previous work on low-mass star instability}
The excitation mechanisms known to operate in solar like and LMS are the stochastic and, also for the latter, the $\epsilon$ mechanism. Stochastic oscillations are driven by turbulent convection and have been detected in solar-like, $\beta$ Cephei and red giant stars, which have large outer convection zones. The $\epsilon$ mechanism is produced by the high sensitivity of certain nuclear reaction rates to temperature, and always has a destabilizing effect. However, no observational evidence has been reported so far of stars pulsating due to this mechanism, despite theoretical predictions. \citet{palla05} found theoretical excitation of the fundamental radial mode in very low mass stars and brown dwarfs (0.02 $\leq$ M/M$\odot$ $\leq$ 0.1) due to D-burning. A non-radial instability study of similar models by \citet{moya11} found the excitation of non-radial modes due to the same mechanism. Observational variability of both type of objects has been reported (see e.g. \citealt{cody09} and references therein), although it has not yet been unequivocally attributed to pulsations. Photometric variability of M dwarfs (\citealt{davenport11}, \citealt{hartman11}, {\bf \citealt{becker11}, \citealt{kowalski09}}) has been usually attributed to flares and chromospheric or atmospheric activity. Recently, the results of an on-going time-series photometric campaign specifically searching for pulsating M dwarfs has been published by \citet{baran11} with no positive detections.

Previous theoretical studies about M dwarfs instability were done by several authors. \citeauthor{gabriel64} (\citeyear{gabriel64}, \citeyear{gabriel67}) proposed for the first time the instability of completely convective stars on the main sequence, based on the evaluation of a vibrational stability coefficient for 0.27 and 0.16~M$\odot$ models. The fundamental radial mode and its first two harmonics were excited when the effects of convection were considered.

\citet{toma72} predicted the instability of the fundamental radial mode for models of 0.2, 0.6, 1.0 and 2.0~M$\odot$ due to D-burning; while \citet{boury73} found instability of 0.5~M$\odot$ models, with effective temperatures $>$ 4\,000~K, due to H-burning reactions.

\citet{opoien74} did a pulsation study of models from 0.085 to 0.5~M$\odot$ on the H main sequence and of 0.012 to 0.5~M$\odot$ on the D-main sequence. They investigated only the fundamental radial mode, which they found to be excited due to an $\epsilon$ mechanism of the corresponding dominant nuclear species. They found pulsation periods between 10 and 41~min for the H-burning models and between 1 to 20~h for the D-burning ones. Due to the long $e$ folding times, they concluded that the models were marginal candidates to show detectable pulsations. \citet{gabriel77} repeated this work with improved physics in the models obtaining pulsation periods in agreement with those previously calculated and $e$-folding times lowered up to a factor of 4. The authors pointed out that in order to assure the instability of the more massive models, the full non-adiabatic equations should be solved.

The work we present here solves for the first time for M dwarf models the complete non-adiabatic pulsation equations for radial and non-radial modes. All the previous works in the literature are adiabatic oscillation studies, some of them only radial, which based their results in the evaluation of a vibrational stability coefficient.

\begin{table}
\caption{Main physical parameters of the excited models. The period ($P_0$) and e-folding time ($\tau_{fold}$) of the fundamental mode ($\ell$=0, $k$=0) are also given. The last column gives $e^{\Delta t/|\sigma_I|}$ for the models excited by D-burning. {\bf See text for details.}} 
\label{tab:t1}
\centering
\renewcommand{\footnoterule}{}
\renewcommand{\tabcolsep}{3pt}
\begin{tabular}{ccccccccc}
\hline
  M          & age   &  T$_{eff}$  & $\log L$ & R   & $\log$ g &  $P_0$   & $\tau_{fold}$ & $e^{\Delta t/|\sigma_I|}$ \\
  ($M\odot$) & (Myr) &    (K)     & ($L\odot$) & ($R\odot$)  &  dex    & (min)    & (Myr)       &      \\
\hline
 0.1  & 0.75      &  3102 & -1.05 & 1.0 & 3.4 &  5.2~h  & 1.23    &  3.1  \\
      & 1.0       &  3107 & -1.07 & 1.0 & 3.4 &  5.1~h  & 1.16    &  3.3  \\
      & 1.5       &  3116 & -1.10 & 1.0 & 3.5 &  4.7~h  & 1.42    &  2.7  \\
      & 2.0       &  3129 & -1.16 & 0.9 & 3.5 &  4.2~h  & 2.63    &  1.7  \\
\hline
 0.2  & 0.5       &  3274 & -0.49 & 1.8 &  3.2 &  8.4~h  & 0.57    &  3.5  \\
      & 1\,10$^4$ &  3383 & -2.26 & 0.2 &  5.1 &  23     & 349     &       \\
\hline
0.25  & 1\,10$^4$ &  3467 & -2.06 & 0.3 &  5.0 &  27     & 239     &      \\
\hline
0.4   & 5\,10$^2$ &  3648 & -1.69 & 0.4  & 4.9 &  28     & 189     &       \\
      & 1\,10$^3$ &  3646 & -1.68 & 0.4  & 4.9 &  35     & 178     &       \\
      & 2\,10$^3$ &  3646 & -1.68 & 0.4  & 4.9 &  35     & 170     &       \\
      & 1\,10$^4$ &  3653 & -1.66 & 0.4  & 4.9 &  36     & 144     &       \\
\hline
 0.5  &5\,10$^2$  &  3814 & -1.45 &  0.4  & 4.9 &  38    & 141     &       \\
      &9\,10$^2$  &  3812 & -1.44 &  0.4  & 4.9 &  39    & 133     &       \\
      &1\,10$^3$  &  3812 & -1.44 &  0.4  & 4.9 &  39    & 131     &       \\
      &2\,10$^3$  &  3811 & -1.44 &  0.4  & 4.9 &  39    & 125     &       \\
      &4\,10$^3$  &  3814 & -1.43 &  0.4  & 4.8 &  39    & 120     &       \\
      &6\,10$^3$  &  3816 & -1.43 &  0.4  & 4.8 &  40    & 117     &       \\
      &8\,10$^3$  &  3820 & -1.42 &  0.4  & 4.8 &  40    & 115     &       \\
      &1\,10$^4$  &  3823 & -1.42 &  0.4  & 4.8 &  40    & 113     &       \\
\hline
\end{tabular}
\end{table}

   \begin{figure*}
     \begin{tabular}{cc}
 \resizebox{0.45\linewidth}{6cm}{\includegraphics[angle=90]{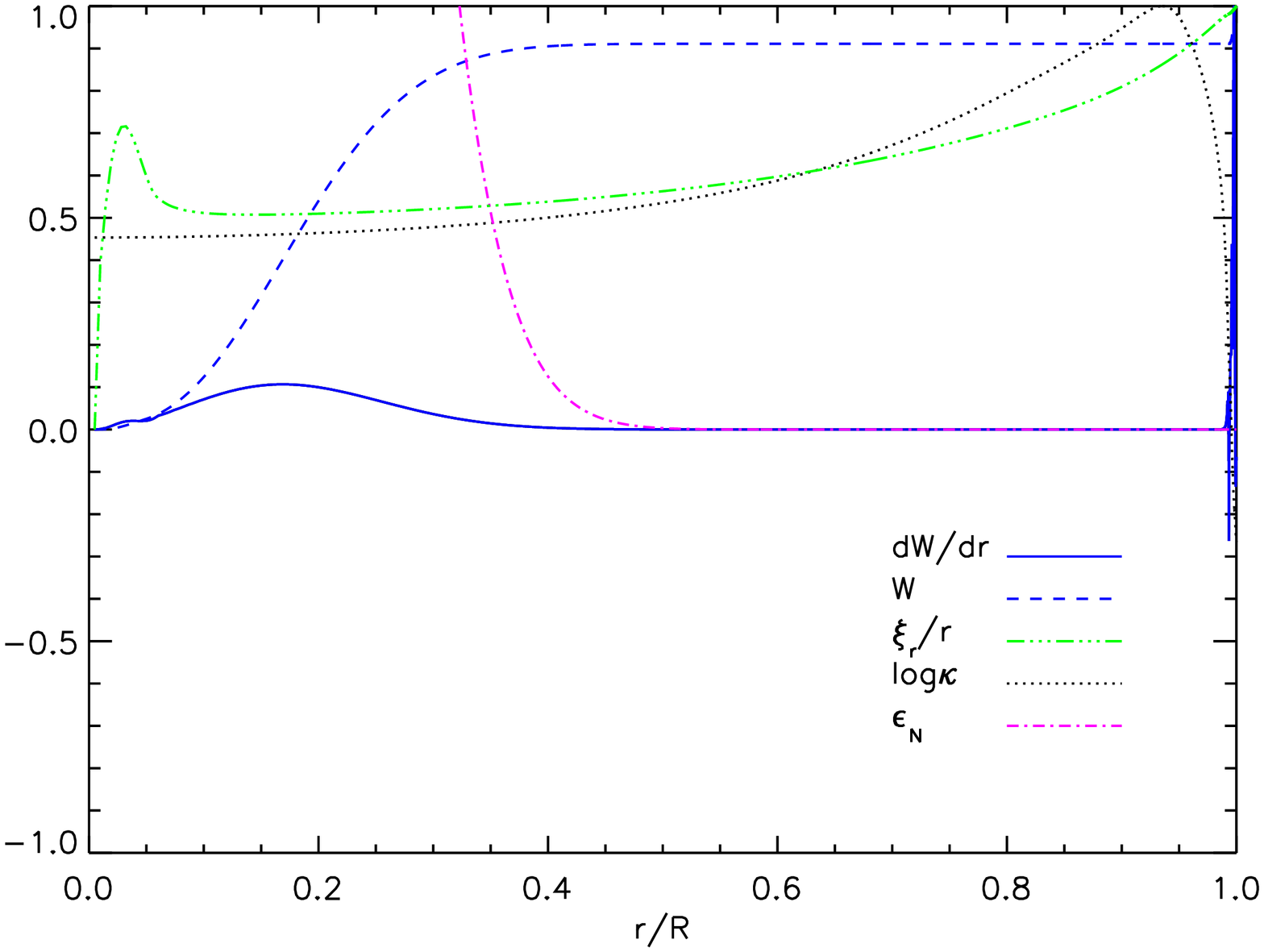}} &
   \resizebox{0.45\linewidth}{6cm}{\includegraphics[angle=90]{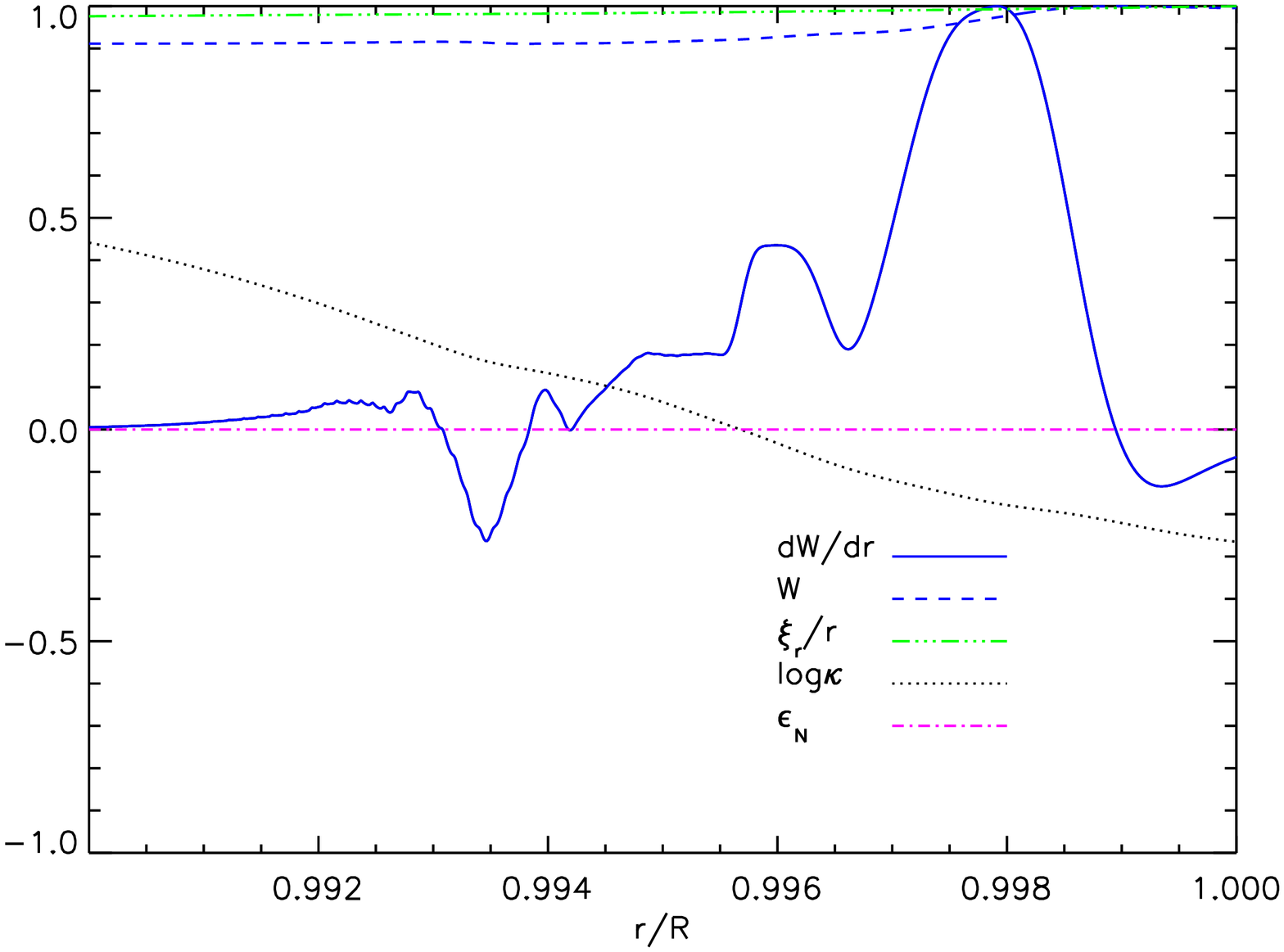}} \\
      \end{tabular}
   \caption{Differential ($dW/dr$) and total work ($W$) done over one oscillation cycle for the 0.1~M$\odot$ model at the peak of maximum D-burning (age 1~Myr). Also shown are the radial displacement ($\xi_r/r$), the opacity ($\log \kappa$) and the nuclear energy generation ($\epsilon_N$). All quantities are normalized to unity except $\epsilon_N$. Right: Zoom of the surface of the model ($r/R=[0.99,1.0]$)}
   \label{fig:model01_l0k0}
   \end{figure*}

\section{Evolutionary Models}
We have used MacDonald's fully implicit stellar evolution code (\citealt{macdonald10} and references therein) to construct evolutionary sequences of model stars of composition X = 0.6958, Y = 0.2855, Z = 0.0187, and masses 0.1, 0.2 0.25, 0.3, 0.4, and 0.5 M$\odot$. Each sequence is begun at a point on the Hayashi contraction phase before the start of D-burning. B\"{o}hm-Vitense convection is used with $\alpha = 1.699$. The reaction network follows the detailed evolution of D- and He$^3$, and includes all reactions in the ppI chain as well as the dominant reactions of the ppII and ppIII chains.

\section{Pulsation Analysis}
We have used the non-radial non-adiabatic oscillations code GraCo (Granada Code, \citealt{moya08}, \citealt{moya04}) to search for pulsations in our M dwarf evolutionary models with ages from prior to the onset of D-burning and up to the age of the Universe. We calculated {\bf radial and non-radial} modes from $\ell$=0 to $\ell$=3 in the frequency range 20~$\mu$Hz up to {\bf the acoustical cut-off frequency for an isothermal atmosphere of each model, given by $\nu_c=c_s/4\pi H_p$, where $c_s$ is the sound speed in the surface of the star and $H_p$ the pressure scale height} (see \citealt{crl10} for a description of the method).

Our results indicate that {\bf only} the fundamental radial mode is excited due to an $\epsilon$ mechanism caused by the nuclear D-burning (H$^2$($p,\gamma$)He$^3$) operating in the young 0.1 and 0.2~M$\odot$ models; the non-equilibrium He$^3$- He$^3$ reactions (ppI and ppII chains) operating in the 0.2 and 0.25~M$\odot$ models with age 10$^4$~Myr; and a flux blocking mechanism in the partially convective 0.4 and 0.5~M$\odot$ models older than 500~Myr. For the radial pulsations excited by D-burning, the periods, which are related to the dynamical time scale of the object, $\tau_{dyn} \sim (G {\bar\rho})^{-1/2}$, are found to be 4.2 to 5.2~h for the 0.1~M$\odot$ and 8.4~h for the 0.2~M$\odot$ models. The periods of the modes excited due to He$^3$ burning and the flux blocking mechanism are in the range 23 to 40~min.

Table~\ref{tab:t1} gives the age and main physical parameters of the excited models and the period and $e$-folding time of the fundamental mode. {\bf Given that the amplitude of the mode cannot be calculated with a linear oscillations code, we refer to the $e$-folding time, defined as $\tau_{fold}=1/|\sigma_I|$, where $\sigma_I$ is the imaginary part of the eigenfrequency, as an estimation of the observability of the mode. As the time-dependence of the mode amplitude is given by $A e^{\sigma_It}$ , where $A$ is the unknown initial amplitude of the mode, the $e$-folding time, $\tau_{fold}$, gives the time needed to increase the initial unknown amplitude a factor of $e$. Therefore, $e$-folding times shorter than the age of the model, or than the time scale of the nuclear reaction, would allow enough time for the amplitude of the mode to be developed, even if the initial amplitude were low. If it were large, then even short $e$-folding time scales could allow for observable amplitudes.}

The D-burning lifetime for the 0.1~M$\odot$ models is 1.4~Myr, defining the D-burning phase as at least 50\% of the star's luminosity provided by D-burning. With $\tau_{fold}$ of the order, or longer, than the age of the model, the 0.1~M$\odot$ models are not favoured for the observational detection of pulsations, unless their initial amplitudes are large. The last column of Table~\ref{tab:t1} gives an estimation of the possibility of the mode being observed through the quantity $e^{\Delta t/|\sigma_I|}$ (\citealt{toma72}, \citealt{moya11}). This gives an estimation of how much the mode grows in amplitude during the time that the model spends in the D-burning phase, given by $\Delta t$. With a maximum value of 3.3 for the 1~Myr model, at which the D-burning reaches its maximum, the probability of being observed is not very much favoured. 

Figure~\ref{fig:model01_l0k0} (left) shows the work integral ($W$) and its derivative ($dW/dr$) for the fundamental mode of a {\bf representative} 0.1~M$\odot$ model.  {\bf The epsilon mechanism associated to the D-burning is responsible for the driving ($dW/dr > 0$), as the work integral in the central region of the star is mostly dominated by the temperature derivative ($\epsilon_T=\partial \epsilon_N/ \partial T|_\rho$) of the nuclear energy generation ($\epsilon_N$), which in our models has a high value, $\epsilon_T \simeq 12$ at temperatures $T \simeq 10^6$~K.} {\bf The epsilon mechanism is strong enough to overcome the damping of the outer regions of the star when $\epsilon_T$ is large and the central density is low, allowing for larger amplitudes ($\xi_r/r$) in the inner region of the star. As the burning of D- proceeds, $\epsilon_T$ decreases from its initial value $\epsilon_T \simeq 12$, at temperatures $T \simeq 10^6$~K, for the excited models to  $\epsilon_T \simeq 6$ for the stable ones.} Figure~\ref{fig:model01_l0k0} (right) zooms in the surface of the star to reveal a small damping ($dW/dr < 0$) and a large driving region. However, they do not contribute significantly to the total work, as $\sim$90\% of the work integral is given by the $\epsilon$ mechanism in the inner regions. {\bf Even if the $dW/dr$ profile in the outer layers seems to be produced by a $\kappa$-mechanism, the maximum driving should then be produced at maximum opacity, i.e. in Figure~\ref{fig:model01_l0k0} $log \kappa_{max}$ at $r/R \simeq$~0.93 or equivalently at $log T \simeq$~4.6, which corresponds to the partial ionization zone of HeII. However, the maximum driving is produced at $r/R \simeq$~0.998 corresponding to $logT \simeq 3.6$, where no partial ionization zone is expected.} A careful analysis of each of the terms in the energy equation shows that is not the $\kappa$ mechanism which plays a role in the destabilization, but the radiative luminosity, which is very model dependent and not trustworthy for these convective or quasi-convective models.

We follow \citet{moya11} to assess that the instability of the modes is really produced by the $\epsilon$ mechanism: checking the invariability of the work integral for different mechanical outer boundary conditions. We apply the criterion that for instability at least 80\% of the work integral must be provided by the $\epsilon$ mechanism. 

   \begin{figure*}
     \begin{tabular}{cc}
 \resizebox{0.45\linewidth}{6cm}{\includegraphics[angle=90]{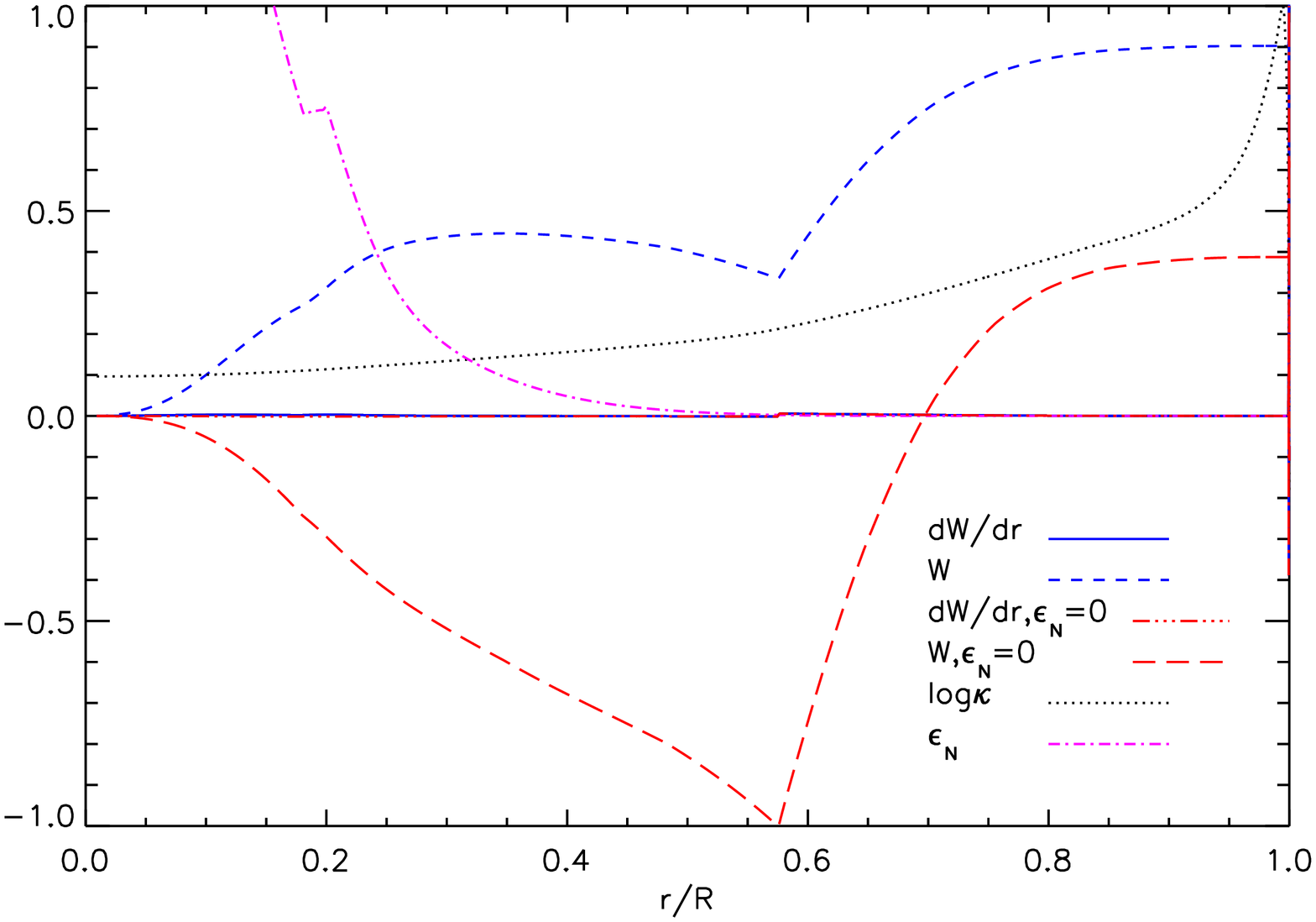}} &
   \resizebox{0.45\linewidth}{6cm}{\includegraphics[angle=90]{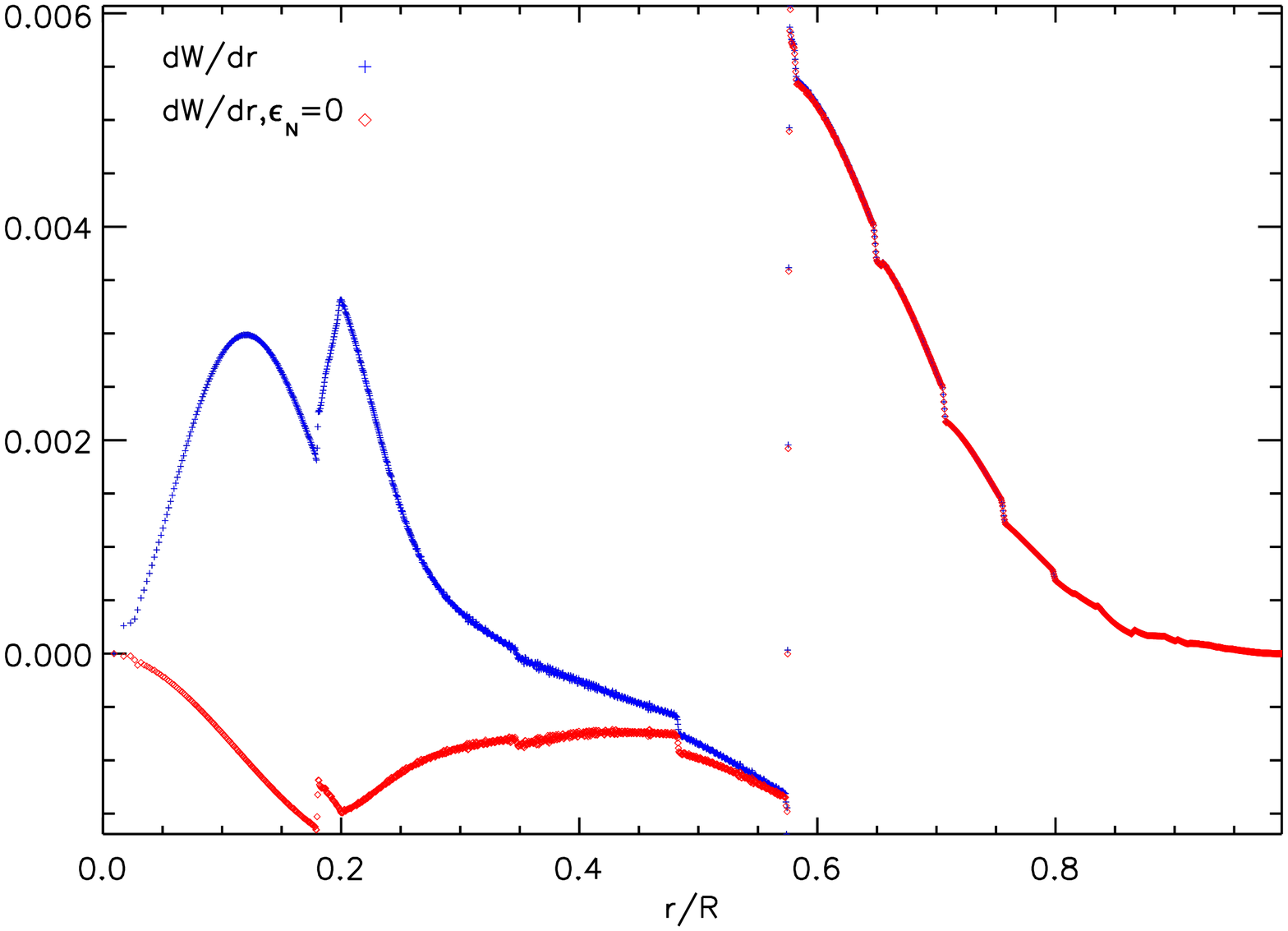}} \\
      \end{tabular}
   \caption{Normalized work (W, blue dashed lines) and its derivative (dW/dr, blue solid lines) for the fundamental radial mode of the 0.5~M$\odot$ model of 500~Myr with and without nuclear reactions ($\epsilon_N$=0, red lines). Right: Zoom of the previous plot in the range $r/R=[0,0.99]$ to see the details of the driving at the base of the outer convection zone, located at $r/R=0.57$ }
   \label{fig:wnucl_nucl0}
   \end{figure*}

The D-burning lifetime decreases with increasing mass, being 0.7~Myr for the 0.2~M$\odot$ models. Only the 0.2~M$\odot$ model at the peak of maximum D-burning (0.5~Myr) is excited via the associated $\epsilon$ mechanism, although with a low probability of being detected, as $\tau_{fold}$ is of the order of the age of the model and consequently  $e^{\Delta t/|\sigma_I|}$ has a low value. The destabilization of the radial fundamental mode for 0.2 and 0.25~M$\odot$ models of 10$^4$~Myr is caused by the $\epsilon$ mechanism powered by the He$^3$ burning {\bf produced in the pp chains.} The period of the modes is 23 and 27~min, respectively, with $e$-folding times low enough for the pulsating amplitude to be developed, although the observational challenge for these elderly objects seems insurmountable.

The 0.4 and 0.5~M$\odot$ models are partially convective: they have an outer, and the younger ones also an inner, convection zone. The excitation of the fundamental radial model for ages older than 500~Myr is due to flux blocking at the base of the outer external convection zone, as described by \citet{guzik00} for $\gamma$ Doradus stars. {\bf The radiative luminosity falls abruptly at the bottom of the convection zone as it can not be transported by radiation. For these models, as the local convective life time at the base of the convection zone is much longer than the pulsation period, the convective flux can not adapt instantaneously to transport the additional energy coming from below. Therefore, this energy is periodically blocked and converted into work driving the oscillations. The large local convective lifetime compared to the pulsation period validates our choice of frozen convection approximation \citep{unno89} in the oscillation code, where variations of the convective flux are not considered during the oscillation}. Fig.\ref{fig:wnucl_nucl0} (left) shows the total work (blue dashed line) and differential work (solid blue line) for the fundamental radial mode of the {\bf representative} 0.5~M$\odot$ model of 500~Myr. The same quantities are shown with red lines for the same model with the nuclear reactions set to zero. The driving region $r/R=[0,0.57]$ is due to the $\epsilon$ mechanism associated to the He$^3$ reactions. When the contribution of the nuclear reactions is switched off, radiative damping is the main contributor to the energy. However, the mode is still excited due to the flux blocking mechanism at the base of the outer convective region at r/R=0.57. This is better seen in Fig.\ref{fig:wnucl_nucl0} (right), where we have left out the r/R=[0.99,1] region, in which the effects of the radiative luminosity on the driving are not trustworthy.

\section{Discussion and Conclusions}
 We presented the first non-adiabatic non-radial pulsational study of M dwarf models. We found the fundamental radial mode to be excited by an $\epsilon$ mechanism caused by D- or He$^3$ burning operating in the fully convective models, or by a flux blocking mechanism for the partially convective ones. The 0.1 and 0.2~M$\odot$ models excited by D-burning have growth time scales of the order of the age of the models, which makes their observational detection elusive, unless their initial amplitudes are large. This contrasts with the results of \citet{palla05} for 0.1~M$\odot$ models of 0.6 and 1.1~Myr which have $e$ folding time scales more favourable for detection.

Oscillations in the 0.2 and 0.25~M$\odot$ of 10$^4$~Myr, excited by the $\epsilon$ mechanism caused by the He$^3$-He$^3$ reactions, are not envisaged to be detected observationally, due to the age and faintness of the models.
 
The observational detection of oscillations for the 0.4 and 0.5~M$\odot$ models excited by a flux blocking mechanism is more likely, as the amplitude of the modes have enough time to develop due to the shorter time scale for the growth of the instability.

{\bf We now discuss whether the predicted pulsation periods could be misinterpreted in observational light curves as rotational periods. \citet{jenkins09} (J09) derive $vsini$ for a sample of 56 M dwarfs with spectral types M3-M6, corresponding to masses in the range 0.1-0.3~M$\odot$, obtaining a distribution that peaks at $vsini=3~km\,s^{-1}$. They add to their data a sample of about 200 M dwarfs with spectral type in the range M0-M9 and $vsini$ calculated in the literature. They found a large spread in $vsini$, from about 0 to 50km/s, for the mid spectral range M4-M6.5 (see figure~1 of J09), although for a large fraction of these stars the measured $vsini$ are upper limits. In general, there is a trend of increasing rotational velocity with later spectral type, i.e. decreasing temperature and mass. We roughly estimate the rotational velocities ($2\pi R/P$) assuming that the pulsation period of our star models is the rotation period (see Table~\ref{tab:t2})}.

\begin{table}
\caption{Estimation of rotational velocities ($2\pi R/P$) for the lowest and highest mass range of the models in our study.}
\label{tab:t2}
\centering
\renewcommand{\footnoterule}{}
\renewcommand{\tabcolsep}{3pt}
\begin{tabular}{ccccc}
\hline
&  M          & R           &  $P_0$   & $v_{rot}$ \\
&  ($M\odot$) & ($R\odot$)  & (min)    & (km\,s$^{-1}$)    \\
\hline
a. & 0.1  &  1  &  5~h  &  243  \\
\hline
b. & 0.2  & 1.8 & 8.5~h &  257  \\
c. &      & 0.2 & 23    &  633  \\
\hline
d. & 0.4  & 0.4 & 23    &  1267 \\
e. &      & 0.4 & 40    &  728  \\
\hline
\end{tabular}
\end{table}

{\bf We conclude that the pulsation periods of the 0.1 and 0.2~$M\odot$ models (corresponding to spectral types about M5-M9) excited by the epsilon mechanism (cases a, b and c) can not be confused with rotation periods, which have maximum values of $vsini < 50~km\,s^{-1}$ (figure~9 of J09). For the highest mass range of our study (0.4-0.5~$M\odot$ models, corresponding to spectral classes about M0-M2) where pulsations are excited by the flux-blocking mechanism, the pulsation periods would be distinguishable from the rotation periods ($vsini<10~km\,s^{-1}$, see again figure~9 of J09).

Furthermore, a M dwarf star that rotates with a period as short as the pulsation periods given in Table~\ref{tab:t1} would likely stand out as a flare star, which would make it an unsuitable target to detect pulsations.}

The observational confirmation of pulsations in M dwarf stars would mean to open the field of asteroseismology to characterize this class of stars with much more precision than it has been done before{\bf , as an independent measure of the stars' mean density could be directly made from the oscillation periods. This would result in a more precise physical characterization of possible planets hosted by pulsating M dwarfs. 

Finally, we can not discard that modes of high radial order in the asymptotic regime are excited by an stochastic mechanism. The study of this possibility, with dedicated codes that include non-linear interactions in the convection zone, would be desirable as it could lead to resolution of the oversizing puzzle for M dwarfs described above.}

\section*{Acknowledgments}

C.R-L acknowledges financial support provided from the {\em Annie Jump Cannon} fund of the Department of Physics and Astronomy of the University of Delaware. A.M. acknowledges the funding of AstroMadrid (CAM S2009/ESP-1496). This research has been partially funded by the Spanish grants ESP2007-65475-C02-02, AYA 2010-21161-C02-02 and CSD2006-00070. We thank P.J. Amado for useful discussions on the manuscript.



\bibliographystyle{aa}
\bibliography{biblio}

\begin{thebibliography}{28}
\expandafter\ifx\csname natexlab\endcsname\relax\def\natexlab#1{#1}\fi

\bibitem[{{Baran} {et~al.}(2011){Baran}, {Winiarski}, {Krzesi{\'n}ski},
  {Fox-Machado}, {Kawaler}, {Dr{\'o}{\.z}dz}, {Faltenbacher}, {Thompson}, \&
  {Reed}}]{baran11}
{Baran}, A.~S., {Winiarski}, M., {Krzesi{\'n}ski}, J., {et~al.} 2011, \actaa,
  61, 37

\bibitem[{{Becker} {et~al.}(2011){Becker}, {Bochanski}, {Hawley}, {Ivezi{\'c}},
  {Kowalski}, {Sesar}, \& {West}}]{becker11}
{Becker}, A.~C., {Bochanski}, J.~J., {Hawley}, S.~L., {et~al.} 2011, \apj, 731,
  17

\bibitem[{{Boury} \& {Noels}(1973)}]{boury73}
{Boury}, A. \& {Noels}, A. 1973, \aap, 24, 255

\bibitem[{{Chabrier} {et~al.}(2007){Chabrier}, {Gallardo}, \&
  {Baraffe}}]{chabrier07}
{Chabrier}, G., {Gallardo}, J., \& {Baraffe}, I. 2007, \aap, 472, L17

\bibitem[{{Cody}(2009)}]{cody09}
{Cody}, A.~M. 2009, in American Institute of Physics Conference Series, ed.
  {J.~A.~Guzik \& P.~A.~Bradley}, Vol. 1170, 630--634

\bibitem[{{Davenport} {et~al.}(2011){Davenport}, {Becker}, {Hawley},
  {Kowalski}, {Sesar}, \& {Cutri}}]{davenport11}
{Davenport}, J.~R.~A., {Becker}, A.~C., {Hawley}, S.~L., {et~al.} 2011,
  astroph.1101.1363

\bibitem[{{Delfosse} {et~al.}(2006){Delfosse}, {Bonfils}, {Forveille},
  {Beuzit}, {Perrier}, {S{\'e}gransan}, {Udry}, {Mayor}, {Bouchy}, {Pepe},
  {Queloz}, \& {Bertaux}}]{delfosse06}
{Delfosse}, X., {Bonfils}, X., {Forveille}, T., {et~al.} 2006, in SF2A-2006,
  ed. {D.~Barret et al.}, 395--398

\bibitem[{{Gabriel}(1964)}]{gabriel64}
{Gabriel}, M. 1964, Annales d'Astrophysique, 27, 141

\bibitem[{{Gabriel}(1967)}]{gabriel67}
{Gabriel}, M. 1967, Annales d'Astrophysique, 30, 745

\bibitem[{{Gabriel} \& {Grossman}(1977)}]{gabriel77}
{Gabriel}, M. \& {Grossman}, A.~S. 1977, \aap, 54, 283

\bibitem[{{Guzik} {et~al.}(2000){Guzik}, {Kaye}, {Bradley}, {Cox}, \&
  {Neuforge}}]{guzik00}
{Guzik}, J.~A., {Kaye}, A.~B., {Bradley}, P.~A., {Cox}, A.~N., \& {Neuforge},
  C. 2000, \apjl, 542, L57

\bibitem[{{Hartman} {et~al.}(2011){Hartman}, {Bakos}, {Noyes}, {Sip{\H o}cz},
  {Kov{\'a}cs}, {Mazeh}, {Shporer}, \& {P{\'a}l}}]{hartman11}
{Hartman}, J.~D., {Bakos}, G.~{\'A}., {Noyes}, R.~W., {et~al.} 2011, \aj, 141,
  166

\bibitem[{{Jenkins} {et~al.}(2009){Jenkins}, {Ramsey}, {Jones}, {Pavlenko},
  {Gallardo}, {Barnes}, \& {Pinfield}}]{jenkins09}
{Jenkins}, J.~S., {Ramsey}, L.~W., {Jones}, H.~R.~A., {et~al.} 2009, \apj, 704,
  975

\bibitem[{{Kowalski} {et~al.}(2009){Kowalski}, {Hawley}, {Hilton}, {Becker},
  {West}, {Bochanski}, \& {Sesar}}]{kowalski09}
{Kowalski}, A.~F., {Hawley}, S.~L., {Hilton}, E.~J., {et~al.} 2009, \aj, 138,
  633

\bibitem[{{L{\'o}pez-Morales}(2007)}]{lopez07}
{L{\'o}pez-Morales}, M. 2007, \apj, 660, 732

\bibitem[{{MacDonald} \& {Mullan}(2010)}]{macdonald10}
{MacDonald}, J. \& {Mullan}, D.~J. 2010, \apj, 723, 1599

\bibitem[{{Morales} {et~al.}(2009){Morales}, {Ribas}, {Jordi}, {Torres},
  {Gallardo}, {Guinan}, {Charbonneau}, {Wolf}, {Latham}, {Anglada-Escud{\'e}},
  {Bradstreet}, {Everett}, {O'Donovan}, {Mandushev}, \& {Mathieu}}]{morales09}
{Morales}, J.~C., {Ribas}, I., {Jordi}, C., {et~al.} 2009, \apj, 691, 1400

\bibitem[{{Moya} {et~al.}(2011){Moya}, {Baraffe}, \& {Barrado}}]{moya11}
{Moya}, A., {Baraffe}, I., \& {Barrado}, D. 2011, submitted

\bibitem[{{Moya} \& {Garrido}(2008)}]{moya08}
{Moya}, A. \& {Garrido}, R. 2008, \apss, 316, 129

\bibitem[{{Moya} {et~al.}(2004){Moya}, {Garrido}, \& {Dupret}}]{moya04}
{Moya}, A., {Garrido}, R., \& {Dupret}, M.~A. 2004, \aap, 414, 1081

\bibitem[{{Mullan} \& {MacDonald}(2001)}]{mullan01}
{Mullan}, D.~J. \& {MacDonald}, J. 2001, \apj, 559, 353

\bibitem[{{Opoien} \& {Grossman}(1974)}]{opoien74}
{Opoien}, J.~W. \& {Grossman}, A.~S. 1974, \aap, 37, 335

\bibitem[{{Palla} \& {Baraffe}(2005)}]{palla05}
{Palla}, F. \& {Baraffe}, I. 2005, \aap, 432, L57

\bibitem[{{Rodr{\'{\i}}guez-L{\'o}pez}
  {et~al.}(2010){Rodr{\'{\i}}guez-L{\'o}pez}, {Moya}, {Garrido}, {MacDonald},
  {Oreiro}, \& {Ulla}}]{crl10}
{Rodr{\'{\i}}guez-L{\'o}pez}, C., {Moya}, A., {Garrido}, R., {et~al.} 2010,
  \mnras, 402, 295

\bibitem[{{Tassoul}(1980)}]{tassoul80}
{Tassoul}, M. 1980, \apjs, 43, 469

\bibitem[{{Toma}(1972)}]{toma72}
{Toma}, E. 1972, \aap, 19, 76

\bibitem[{{Unno} {et~al.}(1989){Unno}, {Osaki}, {Ando}, {Saio}, \&
  {Shibahashi}}]{unno89}
{Unno}, W., {Osaki}, Y., {Ando}, H., {Saio}, H., \& {Shibahashi}, H. 1989,
  {Nonradial oscillations of stars}

\bibitem[{{Yee} \& {Jensen}(2010)}]{yee10}
{Yee}, J.~C. \& {Jensen}, E.~L.~N. 2010, \apj, 711, 303

\end{thebibliography}


\bsp

\label{lastpage}

\end{document}